\documentclass{aa}

\def\la{\;
\raise0.3ex\hbox{$<$\kern-0.75em\raise-1.1ex\hbox{$\sim$}}\; }
\def\ga{\;
\raise0.3ex\hbox{$>$\kern-0.75em\raise-1.1ex\hbox{$\sim$}}\; }

\newcommand{\kms}{km~s$^{-1}\,$}
\newcommand{\ms}{m~s$^{-1}\,$}

\newcommand{\cmm}{cm$^{-3}\,$}
\newcommand{\daa}{$\Delta\alpha/\alpha\,$}
\newcommand{\dmm}{$\Delta\mu/\mu\,$}

\begin{document}

\title{A new approach for
testing variations of fundamental constants over cosmic epochs using FIR
fine-structure lines
}
\author{
S.~A.~Levshakov\inst{1}\thanks{On leave from the Ioffe
Physico-Technical Institute, St. Petersburg, Russia}
\and
D.~Reimers\inst{1}
\and
M.~G.~Kozlov\inst{2,1}
\and
S.~G.~Porsev\inst{2,1}
\and
P.~Molaro\inst{3}
}
\offprints{S.~A.~Levshakov
\protect \\lev@astro.ioffe.rssi.ru}
\institute{
Hamburger Sternwarte, Universit\"at Hamburg,
Gojenbergsweg 112, D-21029 Hamburg, Germany
\and
Petersburg Nuclear Physics Institute, Gatchina, 188300, Russia
\and
Osservatorio Astronomico di Trieste, Via G. B. Tiepolo 11,
34131 Trieste, Italy
}

\date{Received 00  / Accepted 00 }

\abstract{}
{To obtain limits on the variation of the fine-structure constant
$\alpha$ and the electron-to-proton mass ratio $\mu$ over
different cosmological epochs. 
}
{A new approach based on the comparison of redshifts of
far infrared (FIR) fine-structure lines
and low-lying rotational transitions in CO is proposed
which is in principle more sensitive by a factor of $\sim$10 
compared to QSO metal absorption lines.
}
{Estimations of the quotient $F = \alpha^2/\mu$ obtained for two
distant quasars \object{J1148+5251} 
($z = 6.42$) and \object{BR 1202--0725} ($z = 4.69$) 
provide 
${\Delta F}/{F} = (0.1\pm1.0)\times10^{-4}$ and
$(1.4\pm1.5)\times10^{-4}$, respectively.
The obtained limits are consistent with no variation of physical
constants at the level of $\sim$0.01\% over a period of 13 Gyr.  
}
{Upcoming observations of  quasars and 
distant galaxies in FIR fine-structure lines of different species  
and in CO low rotational lines at the  
SOFIA, HSO, and ALMA 
are expected to improve the current limit by, at least, an order
of magnitude. 
}

 
\authorrunning{S. A. Levshakov et al.}
\titlerunning{A new approach for
testing variations of fundamental constants at high-$z$
}
\maketitle

\section{Introduction}

Possible time variation of coupling strengths and particle masses
is now being discussed 
in regard to the accelerating expansion of the Universe. 
Theoretical models imposing extra dimensions predict that dimensionless
quantities like the fine-structure constant, 
$\alpha = {\rm e}^2/(\hbar c) \approx 1/137$, 
the electron-to-proton
mass ratio, $\mu = m_{\rm e}/m_{\rm p}$, etc. depend on the scale length
of extra dimensions in Kaluza-Klein or superstring theories 
(for a review, see 
Garc\'ia-Berro et al. 2007). 
This scale factor may vary with
cosmic time giving rise to variations of fundamental constants 
which are defined in the combined 4D and extra-D space-time. 
Very different patterns from linear and slow-rolling to 
oscillating variations are considered in contemporary theoretical models
(e.g., Marciano 1984; Mota \& Barrow 2004; Fujii 2005).
But until now, the evidence for any changes in fundamental constants has
not been unambiguously asserted.

For example, laboratory measurements do not show 
variations of $\alpha$
at the level of $\dot{\alpha}/\alpha = (-2.6\pm3.9)\times10^{-16}$ yr$^{-1}$
(Peik et al. 2006).
A lower limit can be set by the fission product analysis of 
a natural reactor in Oklo (in units of $10^{-17}$~yr$^{-1}$): 
$-0.6 \leq \dot{\alpha}/\alpha \leq 1.2$  
and
$-4 \leq \dot{\alpha}/\alpha \leq 3$ reported by
Gould et al. (2006) and Petrov et al. (2006), respectively,
which approximately corresponds 
to the redshift $z \simeq$0.14.
At higher redshifts, molecular rotational and atomic resonance
transitions in combination with observations
of the \ion{H}{i} 21 cm hyperfine line were used to constrain
different combinations of physical constants (e.g., Varshalovich \&
Potekhin 1996; Drinkwater et al. 1998; Carilli et al. 2000;
Murphy et al. 2001; Kanekar et al. 2005; Tzanavaris et al. 2007). 
A new constrain on
$\Delta\mu/\mu$ at $z = 0.68$,
\dmm~= $-0.6\pm1.9$ ppm (1$\sigma$ statistical error)
was recently reported by Flambaum \& Kozlov (2007) from the comparison 
of absorption-line spectra of NH$_3$ with  CO, HCO$^+$, 
and HCN in radio frequency region\footnote{Note that 
$\mu = m_{\rm p}/m_{\rm e}$\
in Flambaum \& Kozlov (2007).\\
1 ppm $\equiv 10^{-6}$. }.     
This value of $\Delta\mu/\mu$ being linear extrapolated to $z = 0$ corresponds
to $\dot{\mu}/\mu = (-1\pm3)\times10^{-16}$ yr$^{-1}$.
Here we adopt a cosmology with
$H_0 = 70$ km~s$^{-1}$~Mpc$^{-1}$, $\Omega_\Lambda = 0.7$, 
and $\Omega_{\rm m} = 0.3$
(look-back time $t_z = 6.3$ Gyr at $z = 0.68$).

In optical spectra of QSOs, limits on \daa\ of comparable  accuracy 
can be obtained 
from metal absorption line measurements. 
Special observations aimed at
such measurements were performed at two different spectrographs~-- 
the HARPS at the ESO 3.6-m telescope (Chand et al. 2006) and 
the UVES at the ESO 8-m telescope (Levshakov et al. 2007a).
The following values for \daa\ were obtained 
(Levshakov et al. 2007b): 
$-0.12\pm1.79$ ppm at $z = 1.15$ 
and
$5.66\pm2.67$ ppm at $z = 1.84$ 
($1\sigma$ errors including statistical and systematic parts 
are indicated). 
Assuming linear variation with cosmic time,
one finds
$\dot{\alpha}/\alpha = (-0.1\pm2.1)\times10^{-16}$ yr$^{-1}$ and
$\dot{\alpha}/\alpha = (5.5\pm2.6)\times10^{-16}$ yr$^{-1}$, respectively.

To test possible systematics in \daa\ measurements with the UVES spectrograph,
we performed additional control observations of asteroids (Molaro et al. 2007). 
Asteroids provide
an accurate radial velocity reference at the level of 1 \ms, or 0.03 ppm
in units of \daa. 
Since no systematic shifts larger than 1.5 ppm have been revealed, 
the positive signal at $z = 1.84$
may be real or induced by other yet unknown systematics.

At higher redshifts, $z \ga 3.5$, the changes in fundamental constants
can be probed through observations of distant galaxies and QSOs in
far infrared (FIR) and radio ranges. 
Carbon, nitrogen and oxygen FIR fine-structure
lines were suggested 
to probe the interstellar medium in galaxies at cosmological distances and 
to search for
$z > 10$ objects (Petrosian et al. 1969; Loeb 1993; Stark 1997; Suginohara et al. 1999;
Blain et al. 2000; Boselli et al. 2002; Papadopoulos et al. 2004; Schaeres \& Pell\'o 2005; 
Nagamine et al. 2006). 

In this paper 
we report on the limits on $\alpha^2/\mu$ estimated
for the first time from the 
[\ion{C}{ii}] 158 $\mu$m line combined with CO rotational lines
detected in the spectra of the most distant quasar 
\object{J1148+5251} at a redshift $z = 6.42$ 
(Maiolino et al. 2005)
and in the northern component of a pair of AGN hosts 
\object{BR 1202--0725} at $z = 4.69$ (Iono et al. 2006).
The corresponding look-back time is
$t_z = 12.9$ Gyr and $12.5$ Gyr
which is, respectively, 93\% and 91\% of the age of the Universe.

\section{Computational method}

In the nonrelativistic limit and for infinitely heavy point-like
nucleus all atomic transition frequencies are proportional to
the Rydberg constant, $\mathcal{R}\approx 109737.3$~cm$^{-1}$. In this
approximation, the ratio of any two atomic frequencies 
does not depend on any fundamental constant. Relativistic
effects lead to the corrections to an atomic energy which are
proportional to the product $(\alpha Z)^2\mathcal{R}$, where 
$Z$ is atomic number. 
Corrections accounting for the finite nuclear mass
are proportional to $\mu \mathcal{R}$. 
For atoms these corrections are typically much smaller than
relativistic corrections, but they become important for molecules.

To study the dependence of atomic frequencies on $\alpha$  it is
convenient to expand them near the present-time value of $\alpha$
in the co-moving reference frame:
\begin{equation}\label{EQ3}
\omega_{\rm z} = \omega + q x + \dots, \quad x \equiv
\left({\alpha_{\rm z}}/{\alpha}\right)^2 - 1\, .
\end{equation}
Here $\omega$ and $\omega_z$ are the frequencies corresponding to
the present value of $\alpha$ and to a change 
$\alpha \rightarrow \alpha_z$ presumably at epoch $z$. The parameter $q$
(so-called $q$-factor) is individual for each atomic transition
(Dzuba et al. 2002).  

If $\alpha$ is not a constant, the parameter $x$ differs from zero.
In linear approximation, $|\Delta \alpha/\alpha| \ll 1$,
the corresponding frequency shift, $\Delta\omega = \omega_z - \omega$, is given by:
\begin{equation}\label{EQ4}
{\Delta\omega}/{\omega} =
2 {\cal Q}\,{\Delta\alpha}/{\alpha}\, ,
\end{equation}
where ${\cal Q} = q/\omega$ is the dimensionless sensitivity coefficient,
and $\Delta \alpha = \alpha_z - \alpha$.

For distant objects such a frequency shift 
would cause an apparent change in the redshift
\begin{equation}\label{EQ4a}
 {\Delta\omega}/{\omega} = -\Delta z/(1+z)\, ,
\end{equation}
where $\Delta z = \tilde{z} - z$ is the difference between an apparent redshift
$\tilde{z}$ and its true value $z$. 
If $\omega'$ is the observed frequency from a distant object, 
then the true redshift is given by
 \begin{equation}\label{EQtrue}
 1+z = \omega_z/\omega' \, ,
 \end{equation}
whereas the shifted (apparent) value is
 \begin{equation}\label{EQfalse}
 1+\tilde{z} = \omega/\omega' \, .
 \end{equation}

Relativistic corrections grow with atomic number $Z$, but for
optical and UV transitions in light elements they are small:
${\cal Q} \sim (\alpha Z)^2 \ll 1$. For example, the UV
\ion{Fe}{ii} lines 
used in the \daa\ measurements have sensitivities
$|{\cal Q}| \sim 0.03$ (Porsev et al. 2007).
All other metal lines usually observed in quasar spectra have 
even lower values of ${\cal Q}$ except for
some of the \ion{Fe}{i} or
\ion{Zn}{ii} transitions (Dzuba et al. 2002; Dzuba \& Flambaum 2007),
which are, however, too weak to be suitable for precise measurements
of redshifts from quasar spectra. 

However, in the radio and FIR ranges one can significantly increase
the sensitivity to $\alpha$-variation by
looking at transitions between the fine-structure levels. 
In the nonrelativistic limit ($\alpha \rightarrow 0$) such levels are
exactly degenerated. The corresponding transition frequencies
$\omega_{\rm fs}$ are proportional to $(\alpha Z)^2$ what means that
for these transitions ${\cal Q} = 1$,
i.e. they are about 30 times more sensitive to changes in $\alpha$ than UV lines. 
This gives
\begin{equation}\label{EQ5}
{\Delta\omega_{\rm fs}}/{\omega_{\rm fs}} =
2 {\Delta\alpha}/{\alpha}\, .
\end{equation}

In order to utilize high sensitivity of the fine-structure transitions
in differential measurements of $\alpha$, we need an independent
reference source. 
Good candidates for this purpose are    
low-lying rotational lines of CO: carbon monoxide is the second abundant
molecule after H$_2$, 
it is observed in many extragalactic objects\footnote{Emission lines
from the rotational transitions of CO were detected in $\sim$40 galaxies
at $z > 1$ (for a review, see Solomon \& Vanden Bout 2005), and recently at
$z = 5.0267$ towards SDSS \object{J033829.31+002156.3} (Maiolino et al. 2007)
and $z = 5.7722$ towards SDSS \object{J0927+2001} (Carilli et al. 2007).}, and 
frequencies of the rotational transitions of CO (hundreds GHz) fall close
to the frequency of the [\ion{C}{ii}] 
fine-structure transition $^2$P$_{3/2} \rightarrow ^2$P$_{1/2}$ at 1900.539 GHz. 

In a good approximation,
the frequencies $\omega_{\rm rot}$ of the rotational lines of light
molecules are independent of $\alpha$, but
sensitive to the electron to proton mass ratio $\mu$:
$\omega_{\rm rot} \sim \mu \mathcal{R}$, i.e. 
\begin{equation}\label{EQ6}
{\Delta\omega_{\rm rot}}/{\omega_{\rm rot}} =
{\Delta\mu}/{\mu}\, .
\end{equation}

If $\alpha$ and/or $\mu$
changes in course of cosmic time, 
the apparent redshifts for the fine-structure line(s)
and for the rotational line(s) will 
differ. Using Eqs.~(\ref{EQ4a}), (\ref{EQ5}), and
(\ref{EQ6}), one finds 
\begin{equation}\label{EQ7}
(\tilde{z}_{\rm rot} - \tilde{z}_{\rm fs})/(1 + z) \equiv {\Delta z}/(1 + z) =
2 {\Delta\alpha}/{\alpha} - {\Delta\mu}/{\mu}\, ,
\end{equation}
where $z$ is a reference redshift.

By introducing the parameter $F = \alpha^2/\mu$, we can rewrite Eq.(\ref{EQ7})
in the form
\begin{equation}\label{EQ8}
{\Delta z}/(1+z) = {\Delta F}/{F}\, .
\end{equation}
One can also
keep in mind that variation of $\mu$ is by far larger than that of $\alpha$, as
predicted by some GUT models (e.g. Langacker et al. 2002; Flambaum 2007).

\section{Results}

Now we apply Eq.(\ref{EQ8}) to the observations of [\ion{C}{ii}] 158 $\mu$m
and CO emission lines from the quasars \object{J1148+5251} 
and \object{BR 1202--0725}
to obtain a limit on variation of the parameter $F$. 

The spectral
observations of the [\ion{C}{ii}] 158 $\mu$m emission line in the quasar \object{J1148+5251}
were carried out with the Institut de Radioastronomie Millim\'etrique (IRAM) 30-m telescope 
at the frequency of 256.1753 GHz covering a bandwidth of 1~GHz with 256 channels
spaced by 4~MHz (Maiolino et al. 2005). At this frequency the resolving 
power of the telescope is
9.6 arcsec (Half Power Beam Width) and the 1~GHz bandwidth corresponds to 1170 \kms. 
The resulting spectral resolution and the noise in the 
coadded and rebinned spectrum were, respectively, 56 \kms\ and  0.3 mK (2.8 mJy), 
leading to a 30\% accuracy of the flux density scale.
The [\ion{C}{ii}] line was detected at a significance level of $8\sigma$ for
the total exposure time of 12.4~h.
The redshift and the peak intensity of the [\ion{C}{ii}] line  
are $z_{\rm fs} = 6.4189 \pm 0.0006$ and $I_{158} = 11.8$ mJy. 
The reported error $\sigma_z = 0.0006$ corresponds to the uncertainty of the line
position measurement of $\sigma_v \sim$24 \kms, which is about one bin size in
the [\ion{C}{ii}] spectrum at the Nyquist limit of 2 resolution elements. 

Observations of the CO (7$\rightarrow$6) and (6$\rightarrow$5) emission lines were
obtained with the IRAM Plateau de Bure interferometer at the frequencies 108.724 GHz 
(the total integration time $T_{\rm exp} = 22$ h) and
93.206 GHz ($T_{\rm exp} = 14$ h), 
respectively (Bertoldi et al. 2003). At about 5 arcsec angular resolution
($5.7''\times4.1''$ at 3.2 mm)
the CO emission line is unresolved and coincides within the astrometric uncertainties of
$\pm0.3$ arcsec with the optical position of the quasar given by Fan et al. (2003).
The coadded 3~mm data were rebinned to 64 \kms\ ($J=7\rightarrow6$) and
55 \kms\ ($J=6\rightarrow5$) resulting in the accuracy of the line position
measurements of $\sigma_v \sim$36 and 24 \kms, respectively.
These uncertainties are again of a bin size in the reduced spectra.  
The redshifts and the peak intensities of the CO (7$\rightarrow$6)
and (6$\rightarrow$5) lines are, respectively,
$z^{(7-6)}_{\rm rot} = 6.4192 \pm 0.0009$, $I_{(7-6)} = 2.14$ mJy, and
$z^{(6-5)}_{\rm rot} = 6.4189 \pm 0.0006$, $I_{(6-5)} = 2.45$ mJy.  

Weighting the reported rotational redshifts with these peak intensities, 
one obtains the mean
$z_{\rm rot} = 6.4190 \pm 0.0005$. We will take this value for the
quasar's systemic redshift $z$. 
Using the reported redshift $z_{\rm fs}$ and the averaged $z_{\rm rot}$,   
Eq.(\ref{EQ8}) yields
${\Delta F}/{F} = (0.1 \pm 1.0)\times10^{-4}$.

The second [\ion{C}{ii}] line was detected at $z_{\rm fs} = 4.6908$ 
towards the northern component of the quasar \object{BR 1202--0725} (Iono et al. 2006). 
The profile of this line is similar to the
CO (5$\rightarrow$4) and (7$\rightarrow$6) lines seen at
$z_{\rm rot} = 4.6916$ from the same component (Omont et al. 1999). 

The [\ion{C}{ii}] 158 $\mu$m emission  was observed 
with the Submillimeter Array interferometer (SMA, Ho et al. 2004).
The total exposure time at a redshifted [\ion{C}{ii}] frequency of 333.969 GHz
was $T_{\rm exp} = 19.6$~h, and the angular resolution was $3.4''\times2.7''$.
The coadded spectrum was averaged using 120 \kms\ bin size resulting in
the rms noise of 7.5 mJy, or the signal-to-noise ratio S/N$\sim 3$ (the peak flux
density $\sim$23 mJy as shown in Fig.~1 in Iono et al.). 
Assuming the uncertainty of the [\ion{C}{ii}] line position
as $\sim$1/4 bin size, one gets the error $\sigma_v \simeq 30$ \kms. 

The CO $J=5\rightarrow4$ line observed with the IRAM interferometer 
is detected at the $\sim 5\sigma$ confidence level
($T_{\rm exp} \sim 16$~h), and 
the CO $J=7\rightarrow6$ line observed with the IRAM 30-m telescope~-- 
at the $3\sigma$ confidence level. 
The angular resolution in the interferometric observations was $5.0\times2.5$ arcsec, and  
velocity resolution of about 60 \kms. The redshift of the northern component is
$z_{5-4} = 4.6916$. The uncertainty of this value is, probably, 25-30 \kms, i.e.
approximately one resolution element, considering
rather noisy line profiles shown in Fig.~2 in Omont et al.
The angular resolution of the 30-m telescope for the 2-mm beam is 17 arcsec. 
The error of the reported redshift $z_{7-6} = 4.6915\pm0.001$ 
corresponds to the radial velocity uncertainty of 53 \kms.  

Taking into account that the angular resolutions are similar in observations
of the [\ion{C}{ii}] and CO $J=5\rightarrow4$ emission lines, we can use their
redshifts
$z_{\rm fs} = 4.6908\pm0.0006$ and $z_{\rm rot} = 4.6916\pm0.0006$  
(both errors correspond to $\sigma_v = 30$ \kms) to calculate
${\Delta F}/{F} = (1.4 \pm 1.5)\times10^{-4}$.
   
\section{Discussion}

While comparing the redshifted frequencies of {\it different species} to
measure hypothetical variations of physical constants, one must 
account for random Doppler shifts of the line positions
caused by non-identical spatial distributions
of species (referred to as the Doppler noise hereinafter)
which can mimic non-zero signals in \daa\ or \dmm\ or
in a combination of these quantities (e.g., Levshakov 1994; Carilli et al. 2000;
Bahcall et al. 2004).
To quantify  uncertainties induced by the Doppler noise
a sample of $(\Delta\alpha/\alpha)/(\Delta\mu/\mu)$ measurements 
is to be collected. 

The main problem here is how to estimate 
the dispersion of random velocity shifts $\sigma_{\rm v}$ for 
a given system of spectral lines.
In case of a large sample size the value of $\sigma_{\rm v}$ can be found
from the scatter of points. 
For a single measurement, a guess for $\sigma_{\rm v}$
comes from the comparison with data on velocity differences between
spectral lines of similar species in nearby clouds.
Observations of local galaxies show that the
intensity of [\ion{C}{ii}] is strongly correlated with the intensities of 
the low-lying rotational lines of CO,
the fine-structure lines of [\ion{C}{i}] $\lambda\lambda370, 609$ $\mu$m and
[\ion{O}{i}] $\lambda\lambda63, 146$ $\mu$m 
(Malhotra et al. 2001), and the fine-structure line of
[\ion{N}{ii}] $\lambda205$ $\mu$m (Petuchowski \& Bennett 1993; Abel 2006).
However, the surface distribution of the [\ion{C}{ii}] emission may not precisely
follow the actual $^{12}$CO contours (Stacey et al. 1985).
The CO rotational lines, if optically thin, 
are emitted throughout the whole molecular cloud.  
As for the [\ion{C}{ii}] emission, 
it is usually enhanced at the edges of the molecular cloud 
in the photodissociation regions (PDRs). Additionally,
diffuse gas from the \ion{H}{ii} regions 
can contribute to the intensity of the `PDR' lines (Kaufman et al. 1999). 
However, the impact from the diffuse gas decreases with increasing gas densities and drops 
from $\sim$30\% at $n_{\rm H} \sim 1$ \cmm\ to $\sim$10\% at
$n_{\rm H} \sim 10^3$ \cmm\ (Kaufman et al. 2006).

Because both considered high-redshift [\ion{C}{ii}] emitters belong to the
giant molecular gas complexes with $M \sim 10^{10}M_\odot$ (Omont et al. 1999;
Walter et al. 2004), 
the contribution to the integrated observed intensity of the
[\ion{C}{ii}] emission from the \ion{H}{ii} regions is probably negligible.
In particular, Iono et al. (2006) 
note an excellent agreement between the [\ion{C}{ii}] and CO
profiles at $z = 4.69$. 
At $z = 6.42$, Maiolino et al. (2005) favor $n_{\rm H} \sim 10^5$ \cmm\
to explain the observed luminosities in FIR, CO, and [\ion{C}{ii}] lines.

Our estimate of the value of the Doppler noise
for the CO and [\ion{C}{ii}] emission lines at $z \sim 6$ is based 
on typical parameters
of the interstellar molecular clouds with similar gas densities:
$n = 10^2-10^4$ \cmm, $M = 10^2-10^4M_\odot$, size $L = 2-20$ pc (Table~1 in Mac Low 
\& Klessen 2004).  
Using the velocity dispersion and region size relation,
$\sigma_v$(\kms)~= $1.10 L^{0.38}$(pc), empirically derived for molecular
clouds by Larson (1981) and then studied in detail by Solomon et al. (1987) 
and by other groups (e.g., Falgarone et al. 1992; 
Miesch \& Bally 1994; Caselli \& Myers 1995; Frieswijk et al. 2007),
we find $\sigma_v \sim 1-3$ \kms.
Thus, we may assume that if the [\ion{C}{ii}] and CO lines
arise co-spatially, then the velocity 
offset between them does not exceed the velocity dispersion
within the cloud which is of order of a few \kms. 
On the other hand, 
our derived limit $|\Delta F/F| < 10^{-4}$
corresponds to $\sigma_v = 30$ \kms which is the error
of the line position measurement.
From these values, we are confident that
the quotient $F = \alpha^2/\mu$  remains constant at the level of 0.01\%
to a look-back time of 13 Gyr.
 
The reliability of the result obtained can be compared with previous
radio observations of molecules at intermediate redshifts. 
Kanekar et al. (2005) used 
observations of \ion{H}{i} 21~cm and OH 18~cm 
lines from two absorbers at $z \simeq 0.7$ and found  
$\Delta f/f = (0.44\pm0.36_{\rm stat}\pm1.0_{\rm syst})\times10^{-5}$. 
Here $f = g_{\rm p}(\alpha^2/\mu)^{1.57}$, and $g_{\rm p}$ 
is the proton gyromagnetic ratio.
The systematic error was set from the assumption
that the Doppler noise is 3 \kms. 
However, this seems to be too optimistic taking into account
observations of the nearby diffuse/translucent gas clouds in our
Galaxy. For instance, absorption profiles of the OH 18 cm and HCO$^+$ 3 mm
lines studied at 140-240 \ms\ resolution towards extragalactic compact
continuum sources (Liszt \& Lucas 2000) exhibit, in general, similar
complex structures over the radial velocity interval of $\sim$10-20 \kms,
while showing small differences in details on the scale of 1-2 \kms.
However, comparison of HCO$^+$ absorption with \ion{H}{i} 21 cm 
absorption towards the same targets reveals significant discrepancies.
Figs.~6 and 7 from Liszt \& Lucas show that there are rather wide
velocity intervals ($\sim$20 \kms) where \ion{H}{i} absorption is seen
but HCO$^+$ is absent. It seems quite possible that the Doppler noise
for a sample of \ion{H}{i} 21 cm and molecular lines could be
as large as 10 \kms.
Another set of observations of local warm gas clouds within 15 pc around the Sun
(Redfield \& Linsky 2007) show that
the distribution of cloud velocity difference 
has a width of $\sim$10 \kms. 
This means that in lower resolution observations of quasar
intervening systems the line-of-sight 
velocity differences of order 10 \kms\ may occur on 
a rather short linear scale of $\sim$10-100 pc.

The limit on $\Delta\mu/\mu = -0.6\pm1.9$ ppm
by Flambaum \& Kozlov (2007) was estimated from
the NH$_3$ transitions near $\lambda = 1.26$ cm
and mm-wave lines of CO, HCO$^+$, and HCN detected from
one absorption-line system. 
In this case $\Delta \mu/\mu = 0.289 \Delta v/c$, which means
that the uncertainty of 2 ppm corresponds to $\Delta v  \simeq 2$ \kms.
As discussed above, the systematic error from the Doppler noise 
can be of the same order or even larger. 

To avoid the influence of the Doppler noise on differential measurements of 
physical constants, lines of only one element should be utilized.
For instance, ground state FIR fine-structure transitions 
of [\ion{C}{i}] $\lambda\lambda370, 609$ $\mu$m,
[\ion{O}{i}] $\lambda\lambda63, 146$ $\mu$m,
[\ion{N}{ii}] $\lambda\lambda122, 205$ $\mu$m,
[\ion{O}{iii}] $\lambda\lambda52, 88$ $\mu$m, and 
[\ion{Fe}{ii}] $\lambda\lambda15, 26$ $\mu$m
can be used to directly constrain \daa. 
The sensitivity of this procedure to $\alpha$-variation will be
discussed in the following paper.

Up to now, [\ion{C}{i}] was detected in five high-$z$ 
objects where CO has been observed:
\object{H1413+117}, $J$=1-0 line (Barvainis et al. 1997; 
Wei\ss\ et al. 2005) and
$J$=2-1 line (Wei\ss\ et al. 2003) at $z = 2.557$;
IRAS \object{F10214+4724} and SMM \object{J14011+0252}, 
$J$=1-0 line at $z = 2.285$ and 2.565, respectively (Wei\ss\ et al. 2005);
\object{APM 08279+5255}, $J$=1-0 line at $z = 3.913$ (Wagg et al. 2006); 
\object{PSS 2322+1944}, $J$=1-0 line at $z = 4.120$ (Pety et al. 2004).
Unfortunately, low signal-to-noise ($\la 10$) and 
large uncertainties of 
the positions of [\ion{C}{i}] and CO lines 
($\sim$25 \kms) prevent 
us from constraining relative changes in constants
with an accuracy better than $10^{-4}$.

\section{Conclusions}

In this letter we propose
to use the FIR fine-structure lines of atoms and ions in conjunction with
low-lying rotational lines of CO for probing the cosmic time evolution
of the fundamental physical constants, namely 
the quotient $F = \alpha^2/\mu$. 

The reported constraints on ${\Delta F}/{F}$ are obtained for two
[\ion{C}{ii}] emitters recently discovered at $z = 6.42$ (Maiolino et al. 2005) 
and 
$z = 4.69$ (Iono et al. 2006). We found no evidence for
the variability of $F$ at the level of $10^{-4}$ over a period of 13 Gyr. 
The statistical reliability of this limit seems to be high enough since the expected
value of the Doppler noise for the co-spatially distributed [\ion{C}{ii}] and CO
emission lines, $\sigma_v \sim$ a few \kms, is less that the error of
the line position measurement, $\sigma_v = 30$ \kms.  
However, to be completely confident it should be tested whether
the empirical relationship between
the velocity dispersion and region size derived on base of local measurements is also
applicable to high-$z$ galaxies. 

The result obtained is to be compared with
other tests at extremely high redshifts:
bounds from the height and the positions of the first and subsequent
acoustic peaks of the spectrum of the cosmic microwave background radiation
fluctuations  
($z \sim 1500$, look-back time 13.8 Gyr), 
and  from the relative abundances of light elements
predicted in the big bang nucleosynthesis ($z \sim 10^{10}$) restrict
the fine-structure constant variations at the level of only 2-3\% 
(Ichikawa et al. 2006; Dent et al. 2007; Mosquera et al. 2007).
Thus, the proposed method ensures a significant gain in the accuracy 
and works at the time scales
comparable with the total age of the Universe. 

The limits on ${\Delta F}/{F}$ can be improved by measuring the line
positions with better precision~-- as high as 1-3 \kms\ to be comparable
with the expected Doppler noise.   
The advantage of radio observations for such precise spectral
measurements is twofold: firstly, distant objects inaccessible to
optical observations can be probed, and, secondly, spectral
resolution available in cm- and mm-range 
is much higher than in the optical.
Additionally, the sensitivity to $\alpha$-variation in FIR band is
further increased by roughly a factor of 30 due to larger
${\cal Q}$-factors, as compared to optical band.
In conclusion, the method proposed in this
letter holds the promise of a higher accuracy than obtainable at present
with QSO metal absorption lines.

It is to note, that observations of both
[\ion{C}{ii}] and CO lines in  quasars and
high-redshift galaxies will be one of the key tasks at the   
the Stratospheric Observatory for Infrared Astronomy 
(SOFIA)\footnote{http://www.sofia.usra.edu/},
the Herschel Space Observatory 
(HSO)\footnote{http://herschel.esac.esa.int/}, and
the Atacama Large Millimeter Array 
(ALMA)\footnote{http://www.eso.org/alma/}
which is widely discussed in the literature.
Thus, we can expect that these upcoming observations will
contribute much to clarifying whether fundamental constants vary with cosmic
time or not. 

\begin{acknowledgements}
SAL, MGK, and SGP gratefully acknowledge the hospitality
of Hamburger Sternwarte while visiting there. 
This research has been partly supported by 
the DFG projects SFB 676 Teilprojekt C and RE 353/48-1,
the RFBR grants No. 05-02-16914, 06-02-16489, and 07-02-00210,
and by the Federal Agency for Science and Innovations grant
NSh 9879.2006.2.
\end{acknowledgements}

\end{document}